\documentclass[twocolumn,showpacs,superscriptaddress,prb,floatfix]{revtex4}
\usepackage{amsmath,graphicx,bbm}

\newcommand{\vect}[1]{\mathbf{#1}}
\newcommand{\Det}{\mathrm{Det}}

\begin{document}

\author{Magnus O. Borgh}
\affiliation{Department of Applied Mathematics and Theoretical Physics,
  University of Cambridge, Cambridge, CB3 0WA, UK}
\author{Jonathan Keeling}
\affiliation{Cavendish Laboratory, University of Cambridge,
  J~J~Thomson~Ave., Cambridge CB3 0HE, UK}
\author{Natalia G. Berloff}
\affiliation{Department of Applied Mathematics and Theoretical Physics,
  University of Cambridge, Cambridge, CB3 0WA, UK}

\title{Spatial pattern formation and polarization dynamics of a
  nonequilibrium spinor polariton condensate}

\begin{abstract}
  Quasiparticles in semiconductors --- such as microcavity
  polaritons --- can form condensates in which the steady-state
  density profile is set by the balance of pumping and decay.
  By taking account of the
  polarization degree of freedom for a polariton condensate, and
  considering the effects of an applied magnetic field, we
  theoretically discuss the interplay between polarization dynamics,
  and the spatial structure of the pumped decaying condensate.  If
  spatial structure is neglected, this dynamics has attractors that
  are linearly polarized condensates (fixed points), and
  desynchronized solutions (limit cycles), with a range of
  bistability.  Considering spatial fluctuations about the fixed
  point, the collective spin modes can either be diffusive, linearly
  dispersing, or gapped.  Including spatial structure, interactions
  between the spin components can influence the dynamics of vortices;
  produce stable complexes of vortices and rarefaction pulses with
  both co- and counter-rotating polarizations; and increase the range
  of possible limit cycles for the polarization dynamics, with
  different attractors displaying different spatial structures.
\end{abstract}
\maketitle

\section{Introduction}
\label{sec:introduction}

The experimental realization of quantum condensation of
quasiparticles with finite lifetimes has opened the possibility to
study situations where the steady states of the condensate are not
just controlled by energetics, but involve also consideration of
pumping and decay, leading to steady states with quasiparticle
currents.  The exploration of the possible behavior exhibited in
these systems is driven particularly by recent experimental progress
in microcavity
polaritons\cite{kasprzak06:nature,snoke06:condensation,lagoudakis08,utsunomiya08,amo09:bullet,amo09:sf}
and magnons\cite{demokritov06,demidov08,chumak09,dzyapko09} (as well
as more exotic realizations in helium\cite{volovik08}).  Microcavity
polaritons are the quasiparticles that result from strong coupling
between photons confined in planar semiconductor microcavities and
excitons confined in quantum wells.  The photons are confined by Bragg
reflectors to a two-dimensional cavity, and for small in-plane
momentum have an effectively quadratic dispersion, with a mass
of the order 
of $10^{-5}$ -- $10^{-4}$ of the free electron mass.  From strong
coupling between these photon states and exciton states in quantum
wells, new normal modes arise: polaritons. These are hybrid
light--matter particles, which have both a light
effective mass, as well 
as sufficient interactions to allow a quasi-equilibrium particle
distribution to be established.  However, the distribution is only
quasi-equilibrium, since the photon component is not perfectly
confined, so polaritons have a finite lifetime, and to maintain a
steady state population, continuous pumping is required.  Even when
the energy distribution of polaritons looks reasonably thermal, this
does not mean effects of finite lifetime may be neglected, as one may
have thermalized energy distributions on top of steady state flows
driven by the pumping and decay.

Experimentally, it is observed that above a threshold pumping
strength, an accumulation of low energy polaritons is accompanied
by: a significant increase in temporal coherence, 
spatial coherence that extends over the entire cloud of
polaritons\cite{kasprzak06:nature,snoke06:condensation}, the
appearance of quantized vortices\cite{lagoudakis08}, suggestions of
changes to the excitation spectrum\cite{utsunomiya08}, and a
robustness of polariton propagation to
disorder\cite{amo09:bullet,amo09:sf}.  Of these, the observation
of spontaneous vortices\cite{lagoudakis08} in polariton condensates
provides a strong hint that the steady states of the system are
affected by the flow of particles.  For a more general introduction to
microcavity polaritons, their condensation, and the physical systems
studied there exist several reviews or books,
e.g., Refs.~\onlinecite{ciuti03,keeling07:review,kavokin:oup}.

The aim of this article is to investigate the rich varieties of
behavior that come from combining spatial profiles driven by particle
flow with the dynamics due to the spin degree of freedom that
microcavity polaritons also possess.  The effects of this spin degree
of freedom have been previously considered in equilibrium: due to the
weak attractive interaction between opposite spin polarizations, a
condensate is expected to be linearly
polarized\cite{laussy06,shelykh06}, i.e., to have an equal density of
left- and right-circularly polarized components.  Applying a magnetic
field can then lead to a phase transition, as the density of one
circular polarization increases, and the other decreases, leading
through an elliptically polarized phase to a single circularly
polarized condensate\cite{rubo06,keeling08:spin}.  In the absence of
further symmetry breaking, the linear or elliptically polarized phase
has two gapless modes, corresponding to two broken symmetries:
the overall polariton phase, and the direction of linear polarization
(i.e., orientation of elliptical polarization).  The transition to a
circular polarized state is thus a phase transition, as such a phase
has only a single broken symmetry, and a single gapless mode.

Assuming no other symmetry breaking terms, the linearly polarized
condensate can have independent vortices of the left- and
right-circularly polarized components\cite{rubo07}; such
polarization-dependent phase winding has recently been 
observed\cite{lagoudakis09:half-vortices}.  These two vortices are not
completely independent. Even in the absence of any further symmetry
breaking, there is a short-range density-density
interaction between 
vortices of opposite polarizations\cite{rubo07,keeling08:spin},
however such a term does not depend on the circulations of the two
vortices.  If one also takes account of TE-TM splitting of the
electromagnetic modes\cite{panzarini99}, this leads to a term
that splits linear polarization states according to whether the
polarization is parallel or perpendicular to the polariton momentum;
as such, this provides a circulation dependent interaction between
vortices of the two polarizations\cite{toledo09}.  In addition, real
materials are expected to possess a small splitting between linear
polarization states due to asymmetry of the quantum-well interfaces in
non-centro-symmetric crystals\cite{aleiner92}; such a splitting can
also be controlled by applying electrostatic fields\cite{malpuech06}
or stress\cite{snoke09:stress}.  Such terms will again induce
interactions between vortices of the left- and right-circular
polarization, and if strong enough, will lead to a pinning of the
polarization of a polariton condensate, as has been observed in
experiment \cite{amo05,klopotowski06,kasprzak06:nature,kasprzak07}.

Considering the splitting of linear polarizations as a phase-locking
term between the two circular polarization components, and combining
this with a magnetic field that favors one or other circular
polarization, one has a Josephson problem\cite{leggett01}, where the
energy favoring linear polarization is a Josephson coupling, and the
magnetic field leads to a bias between the two fields.
\citet{shelykh08} have considered the interplay between these spinor
Josephson oscillations, and Josephson coupling between two different
spatial modes of a trapped polariton condensate, showing that
complex behavior can arise in this four-mode system without
  pumping and decay.  A similar problem has also been studied
  in the context of spinor condensates of cold atoms in double well
  potentials\cite{satija09,julia-diaz09}. If one does include pumping
  and decay then even the two-mode system (i.e., just  the
  dynamics of the spin component) can show a rich variety of behavior
  \cite{wouters08:prb,eastham08}.  Due to dissipation, the dynamics
settles on an attractor, which may either correspond to a phase-locked
(i.e., linearly polarized) state, or may correspond to a limit cycle,
in which the phase difference between the two polarization components
continually increases, with the cyclic nature arising from $2\pi$
periodicity in the phase difference.  There also exist parameter
ranges where there is bistability, so that which attractor the system
settles on depends on the initial conditions.  In this article, we
will both review this two-mode dynamics, and then extend to the case
of including many spatial modes in addition to the spin dynamics.


To describe multiple spatial modes in a trapped, pumped, decaying
condensate, a mean-field approach to this problem leads to a complex
Gross--Pitaevskii equation\cite{keeling08:gpe,wouters08:bec}.  It is
also possible to include fluctuations beyond mean-field theory within
such a formalism by quantum stochastic approaches\cite{wouters09}.  As
well as describing the spatial structure of nonequilibrium
condensates, the complex Gross--Pitaevskii equation has also been
applied in a wide range of areas, see
for example Refs.~\onlinecite{aranson02,staliunas03}.  One
application which is 
very closely connected to microcavity polaritons is the dynamics of
lasers propagating through strongly nonlinear materials, where
combinations of cubic and quintic nonlinearities can produce a state
with a preferred density, sometimes referred to as ``liquid
light''\cite{michinel02,michinel06}.  The results in this article
suggest that consideration of polarization dynamics in such materials
may also provide interesting results.

By considering the spinor complex Gross--Pitaevskii equation to
describe the spinor condensate in an harmonic trap, several new
classes of attractors are seen to occur in addition to those present
for the two-mode system.  There are limit cycles describing small
oscillations of the phase difference between the two components, and
limit cycles with $4\pi$ periodicity of the phase difference.  In
addition, the different dynamical attractors correspond to different
spatial profiles, in which the presence or absence of vortex lattices
can be influenced by the applied magnetic fields.

As well as describing the steady state profiles, the complex
Gross--Pitaevskii equation can describe the normal modes for small
fluctuations about such steady states.  In the absence of a spatial
trap, one notable consequence of including pumping and decay is that
the long-wavelength modes of a dissipative condensate are modified, to
become diffusive\cite{wouters07:bec,szymanska06:prl,szymanska07};
similar results are seen both for incoherent pumped condensates, and
for optical parametric oscillation, where scattering of pumped
polaritons into signal and idler states is modified by bosonic
enhancement due to the population of the
signal\cite{wouters05:opo,wouters07:opo}.  From this standpoint,
another purpose of this article is to address how the combination of
pumping, decay, and symmetry-breaking terms modify the
spectrum of 
the spinor condensate.  While a diffusive mode for the overall
polariton density and phase always exists, the long-wavelength modes
for spin modes can show a range of behaviors: diffusive, linearly
dispersing, or gapped, dependent on the balance of decay and
symmetry-breaking terms.

The remainder of this article is arranged as follows.  In Section II
we introduce the model of spinor condensates as a system of coupled
Gross--Pitaevskii equations with pumping and decay terms. By neglecting
the spatial variations of polarized components we discuss the
bifurcation diagrams for a two-mode system in Section III. The normal
modes of the homogeneous system are obtained in Section IV. In Section V
we study the stability of cross-polarized vortices. The detailed study
of the dynamics of the full trapped system is performed in Section
VI. In Section VII we discuss how various regimes can be detected in
experiments. The conclusions in Section VIII summarize our findings.

\section{Model for spinor polaritons}
\label{sec:model-spin-polar}

The model we consider in this article consists of the complex
Gross-Pitaevksii equation (cGPE) as described in
Ref.~\onlinecite{keeling08:gpe}, taking into account the two possible
polarizations of polaritons, written in the basis of left- and
right-circular polarized states, denoted by $\psi_{L,R}$.  In addition
to the terms discussed in Ref.~\onlinecite{keeling08:gpe}, three new
parameters arise from considering the spin degree of freedom in an
applied magnetic field.  Firstly, in addition to an interaction with
the total polariton density $H_{U_0} = (U_0/2) (|\psi_L|^2 +
|\psi_R|^2)^2$, there is an attractive interaction between opposite
spin species, $H_{U_1} = - 2U_1 |\psi_L|^2 |\psi_R|^2$.  Secondly,
there is a magnetic field term $H_{\Omega_B} = (\Omega_B/2) (
|\psi_L|^2 - |\psi_R|^2)$, and finally there is a symmetry breaking
term $H_{J_1} = J_1 (\psi_L \psi_R^\dagger + \text{H.c.})$, which may
naturally arise\cite{kasprzak06:nature,kasprzak07} due to asymmetry at
the quantum-well interfaces\cite{aleiner92}, or may be induced
by electric fields\cite{malpuech06} or stress\cite{snoke09:stress}.
Put together, these yield the coupled cGPE:
\begin{multline}
  \label{eq:1}
  i \partial_t \psi_L
  =
  \left[
    - \frac{\hbar^2 \nabla^2}{2m} + V(r) 
    + U_0|\psi_L|^2 
    + (U_0 - 2U_1) |\psi_R|^2
  \right.\\\left.
    + \frac{\Omega_B}{2}
    + i 
    \left( \gamma_{\text{eff}} - \kappa - \Gamma |\psi_L|^2 \right)
  \right]
  \psi_L
  + 
  J_1 \psi_R,
\end{multline}
\noindent
with the analogous equation for $\psi_R$ following by replacing
$\psi_L \leftrightarrow \psi_R$ and $\Omega_B \to - \Omega_B$.  In
this article, we will either neglect $V(r)$ (in
sections~\ref{sec:review-two-mode}, \ref{sec:norm-modes-extend}), or
consider a harmonic trap, $V(r) = m \omega_0^2 r^2/2$.  In either
case, we will rescale energies and lengths in terms of the
harmonic-oscillator energy $\hbar \omega_0$, and length $l$ such
that $\hbar 
\omega_0 = \hbar^2 / 2m l^2$. By additionally rescaling the density
such that $U |\psi|^2 \to \hbar \omega_0 |\psi|^2$ the cGPE can be
written as:
\begin{multline}
  \label{eq:2}
  i \partial_t \psi_L
  =
  \bigg[
    - \nabla^2 + v(r)
    + |\psi_L|^2 
    + (1 - u_a) |\psi_R|^2
    + \frac{\Delta}{2}\\
    + i 
    \left( \alpha - \sigma |\psi_L|^2 \right)
  \bigg]
  \psi_L
  + 
  J \psi_R
\end{multline}
with the definitions $ \alpha=(\gamma_{\text{eff}} -
\kappa)/\hbar \omega_0,\sigma = \Gamma/U_0, J = J_1/\hbar \omega_0,
\Delta = \Omega_B/\hbar \omega_0, u_a=2 U_1/U_0$. If considering a
harmonic trap, then $v(r) = V(l r)/\hbar \omega_0 = r^2$, otherwise we
will take $v(r) = 0$.

The dimensionless parameter $u_a$ describes the extent of spin
anisotropy of the interaction; using the estimate in
Ref.~\onlinecite{renucci05} we take $u_a=1.1$.  From estimates of the
polarization splitting in Refs.~\onlinecite{klopotowski06,kasprzak07},
we take $J_1 \simeq 0.1 \text{--} 0.2 \text{meV}$.  Using the
estimates of trap frequency given in Ref.~\onlinecite{keeling08:gpe},
we will take $0 \leq \alpha \leq 10, \sigma=0.3$, and we may write the
dimensionless splitting $J \simeq 1$; in the numerical results, we
will consider a variety of values of $J$ around this value.

The spatially extended, harmonically trapped system is unstable unless
pumping is restricted to a finite spot\cite{keeling08:gpe}. For our
studies of this system, we therefore consider a circular pumping spot
with a cutoff radius $r_0$:
$\alpha\rightarrow\alpha\Theta(r_0-r)$. For constant pumping strength,
$r_0$ determines whether the system forms vortices
spontaneously\cite{keeling08:gpe}.

\section{Review of the two-mode model}
\label{sec:review-two-mode}

In order to provide a basis from which to understand the behavior
observed in the trapped system, it is first useful to review the
results that occur in the ``two-mode model'', where spatial variation
of each component is neglected, so the coupled complex
Gross--Pitaevskii equation reduces to two equations for the variables
$\psi_L, \psi_R$.  This model was studied by \citet{wouters08:prb},
where the conditions for the existence of a steady state
(i.e., synchronized) solution were found.\footnote{The results of
  Ref.~\onlinecite{wouters08:prb} were for a slight different
  pumping nonlinearity, but the effects of this difference are not
  significant for steady state, or slowly varying situations.}  As
$\Delta$ increased, \citet{wouters08:prb} found that a synchronized
solution could persist for some range of $\Delta$, with an increasing
phase and density difference between the two modes, until at some
critical $\Delta$ the steady state vanished. Near the critical
$\Delta$, the existence of bistability was noted.  This section will
both summarize these previous results, as well as analyze the
behavior of the desynchronized solution, showing that the dynamics
is analogous to that of the damped driven pendulum, or current-biased
Josephson junction\cite{strogatz94}.

To discuss the dynamics, it is convenient to reparametrize using:
\begin{displaymath}
  \label{eq:3}
  \psi_{L,R} = \sqrt{\rho_{L,R}} e^{i (\phi \pm \theta/2)},
  \quad R=\frac{\rho_L+ \rho_R}{2}, 
  \quad z=\frac{\rho_L- \rho_R}{2},
\end{displaymath}
and write coupled equations for $\theta,R,z$.  (The global phase
$\phi$ does not affect the dynamics.)  In visualizing the dynamics,
one may consider a Bloch vector, defined by $x=\sqrt{\rho_L\rho_R}
\cos(\theta), y=\sqrt{\rho_L \rho_R} \sin(\theta), z$, in terms of the
above variables.  Since there is pumping and decay, the length of the
Bloch vector is not conserved, hence there is a dynamical equation for
$R = \sqrt{x^2+y^2+z^2}$.  The coupled equations have the form:
\begin{align}
  \label{eq:4}
  \dot{\theta} &= - \Delta - 2 u_a z 
  + \frac{2 J z \cos(\theta)}{\sqrt{R^2 - z^2}}
  \\
  \label{eq:5}
  \dot{z} &= 2 (\alpha - 2 \sigma R) z - 2J \sqrt{R^2 - z^2}
  \sin(\theta)
  \\
  \label{eq:6}
  \dot{R} &= 2 \sigma \left( \frac{\alpha}{\sigma} R - R^2 - z^2
  \right).
\end{align}
Writing the equations in this form firstly allows one to understand
the basic effects of pumping and decay on the dynamics, and, secondly,
will be  
shown below to reduce to the damped driven pendulum in a relevant
limit.

\begin{figure}[htpb]
  \centering
  \includegraphics[width=3.2in]{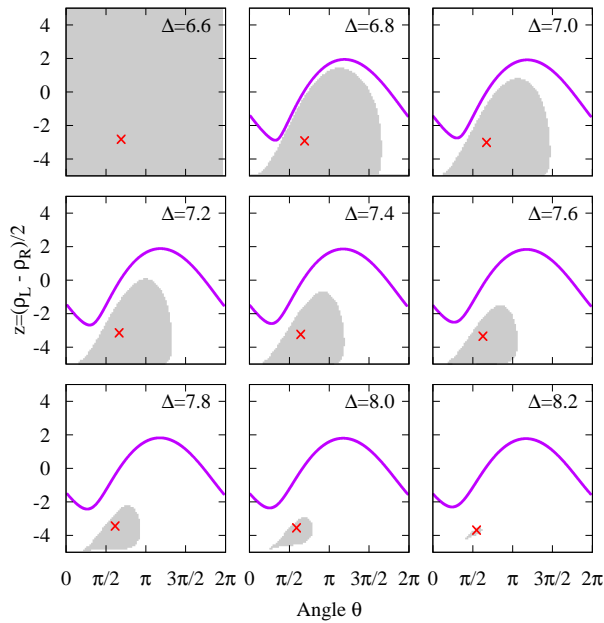}
  \caption{(Color online) Basin of attraction of fixed point and limit
    cycle for full 
    two-mode Josephson problem.  Each panel is for a given value of
    $\Delta$, as indicated, and shows a colormap  according to
    the final state found by starting from the given initial
    conditions $\theta, z$.  Points that flow to the fixed point are
    colored gray, others are white.  The fixed point is marked by a
    red cross, and the limit cycle (when it exists) by the magenta
    line.}
  \label{fig:attraction-basin}
\end{figure}

For a steady state, it is clear that the condition $\dot{R}=0$
defines a Bloch surface, $R(z)$.  One may then study the dynamics of
points that start on this surface to characterize the nature of the
transition as $\Delta$ increases.  As shown in
Fig.~\ref{fig:attraction-basin}, each starting point $z, \theta$ is
attracted either to a fixed point, or to a limit cycle.  The fixed
point is the synchronized solution; the limit cycle is the
desynchronized solution, in which the phase difference $\theta$
between the two spin components is changing --- this behavior is a
cycle since the dynamics are periodic under $\theta \to \theta +
2\pi$.  As $\Delta$ increases, there is a lower critical
$\Delta_{c,\text{lower}}$ below which the all initial conditions on
the Bloch surface flow to the synchronized solution, and an upper
critical $\Delta_{c,\text{upper}}$ which corresponds to the previously
found\cite{wouters08:prb} point beyond which synchronized solutions
can no longer exist.

\begin{figure}[htpb]
  \centering
  \includegraphics[width=3.2in]{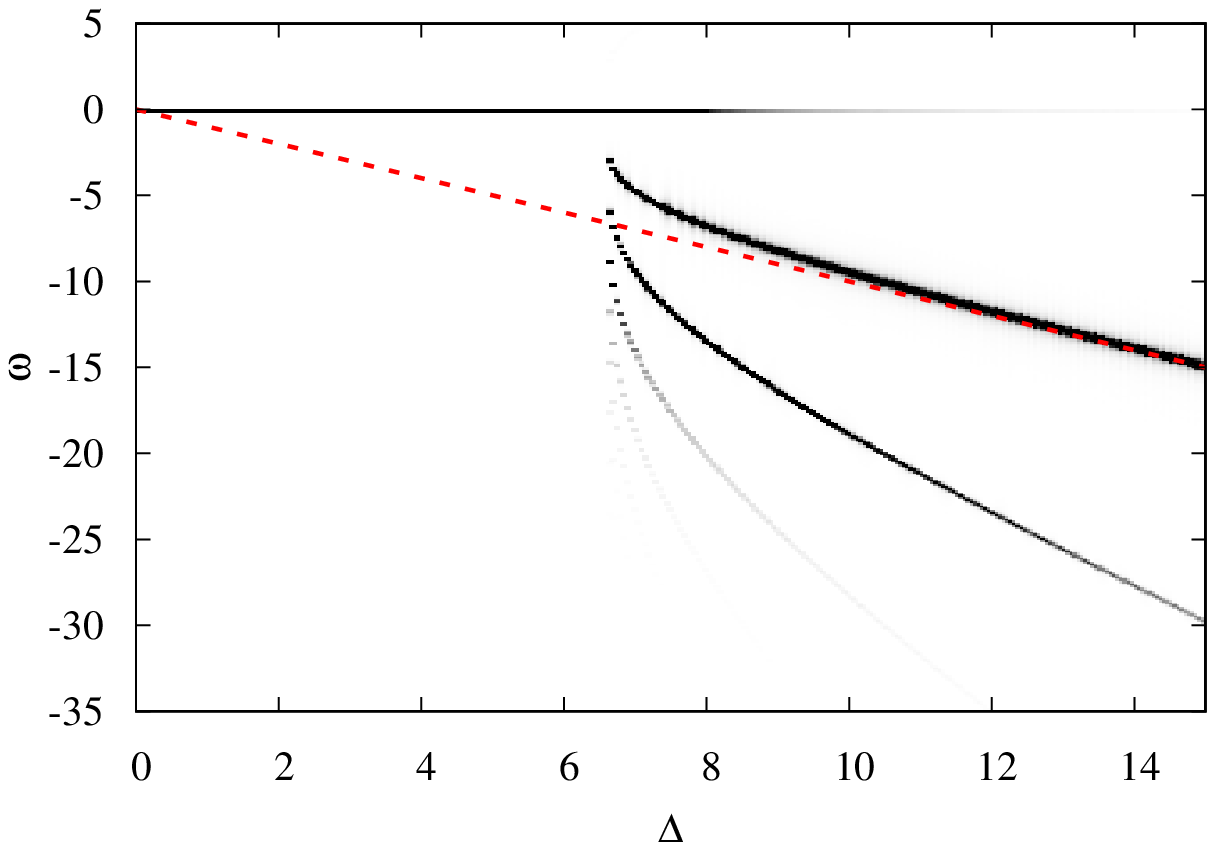}
  \hspace{-2.9in}
  \raisebox{0.35in}{%
  \includegraphics[width=1.4in]{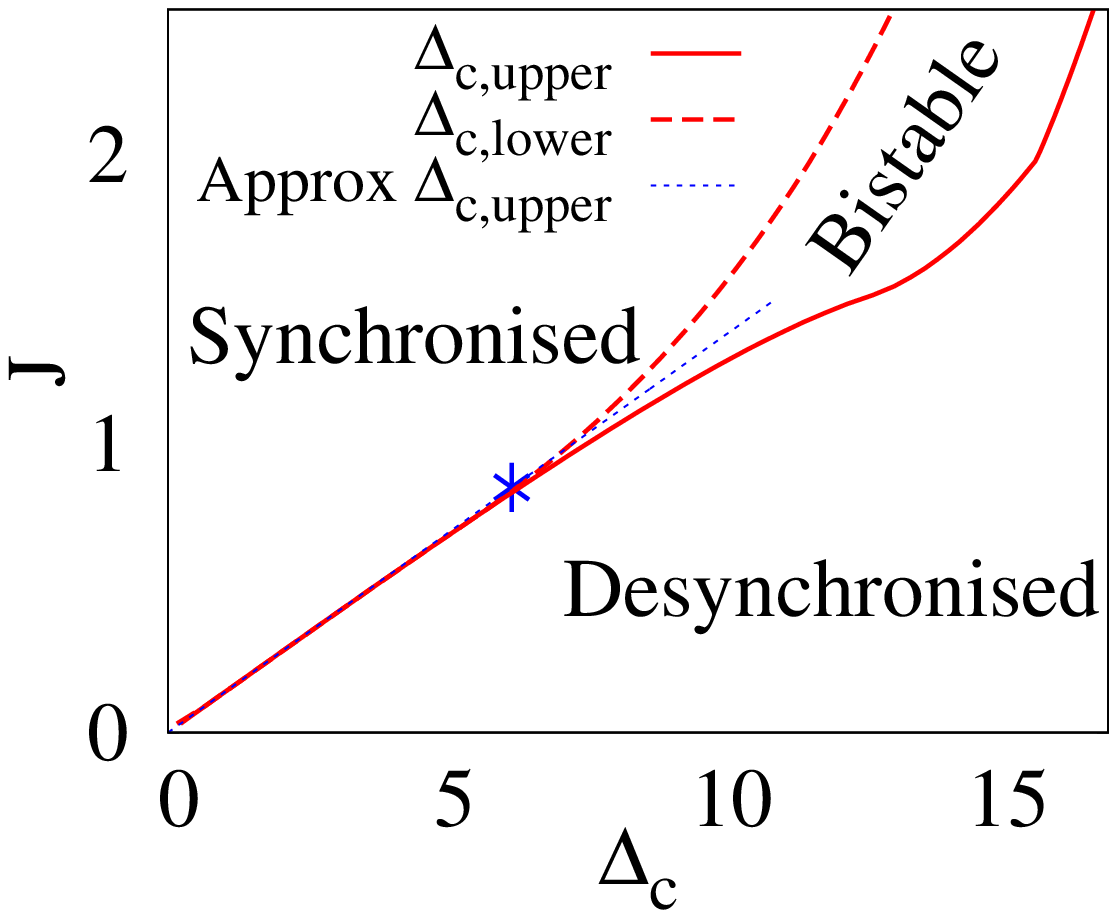}}
  \hspace{1.4in}
  \caption{(Color online) Fourier transform of $\psi_L(t)$, showing
    bifurcation of 
    frequencies as $\Delta$ is increased, and illustrating critical
    behavior $\omega \sim 1/\ln|\Delta-\Delta_c|$.  Initial
    conditions for each $\Delta$ are chosen so that a limit cycle
    (i.e., desynchronized solution) is found for all
    $\Delta>\Delta_{c,\text{lower}}$.  Inset: Regions of stability of
    the synchronized and desynchronized solutions found from
    Eq.~(\ref{eq:4}--\ref{eq:6}).  The dotted line marks the
    approximate $\Delta_{c,\text{upper}}$ appropriate in the Josephson
    regime, Eq.~(\ref{eq:7}), and the $\ast$ marks the known
    bifurcation point\cite{urabe55} in that regime.}  
  \label{fig:ft-two-mode}
\end{figure}

The behavior seen in Fig.~\ref{fig:attraction-basin} can be
understood by noticing that the choice of parameters used puts one in
the ``Josephson regime'' of the Josephson equations\cite{leggett01},
i.e.,  $u_a R \gg J$, and that in addition the typical dynamics obey $z
\ll R$.  In this case, Eq.~(\ref{eq:6}) simply reduces to
$R=\alpha/\sigma$.  Then, by eliminating $z$ from
Eq.~(\ref{eq:4},\ref{eq:5}), one finds:
\begin{equation}
  \label{eq:7}
  \ddot{\theta} + 2 \alpha \dot{\theta} 
  =
  - 2 \alpha \Delta +  4 u_a J \frac{\alpha}{\sigma} \sin(\theta).
\end{equation}
This equation describes a damped driven pendulum, or alternatively a
current-biased Josephson junction.  (N.B., due to our choice of sign of
$J$, ``gravity'' for the pendulum acts to drive $\theta \to \pi$).
The behavior of Eq.~(\ref{eq:7}) is well known (see,
e.g., Ref.~\onlinecite{strogatz94}); for large damping $\alpha$ there
is a simple transition at $ \Delta_{c,\text{upper}} = 2 u_A J/\sigma$
between fixed points for $\Delta< \Delta_{c,\text{upper}}$ and limit
cycles above.  For smaller damping (explicitly, for for $\alpha <
0.595 \sqrt{4 u_a J \alpha/\sigma}$), there is a range of bistability,
where both limit cycles and fixed points may be found\cite{urabe55}.
The inset of Fig.~\ref{fig:ft-two-mode} shows the bifurcation diagram
determined from Eq.~(\ref{eq:4}--\ref{eq:6}), and for comparison, the
value $\Delta_{c,\text{upper}}$ and the critical damping that result
from the approximations leading to Eq.~(\ref{eq:7}).

As well as explaining the classes of behavior seen, Eq.~(\ref{eq:7})
may be used to explain the critical behavior near
$\Delta_{c,\text{lower}}$.  For $\alpha$ sufficiently small that
bistability exists, then near $\Delta_{c,\text{lower}}$, the period of
the limit cycle grows like $T \propto |\ln[\Delta -
\Delta_{c,\text{lower}}]|$.  In the context of the current problem,
this period relates to the average chemical potential difference
between the two components, $\langle \dot{\theta} \rangle = \langle
\mu_L - \mu_R \rangle = 2\pi/T$. For large $\Delta$, one eventually
reaches $\mu_L - \mu_R=\Delta$.  The evolution of $\langle \mu_L -
\mu_R \rangle$ is illustrated in Fig.~\ref{fig:ft-two-mode}, which
plots the spectral weight $|\tilde{\psi}_L(\omega)|^2$, choosing
initial conditions so the limit cycle is obtained for all $\Delta >
\Delta_{c, \text{lower}}$.

\section{Normal modes in the extended homogeneous system}
\label{sec:norm-modes-extend}

Before considering the interplay of spin dynamics with the spatial
profile of a trapped condensate, one may first consider a simpler
problem involving the interplay of spin and spatial dynamics --- the
finite momentum normal modes of the pumped, decaying spinor
condensate.  The normal modes of the spinor condensate without pumping
and decay were discussed in \cite{rubo06,keeling08:spin}.  In the
equilibrium case, for small $|\Delta|$, there is an elliptical
condensate (i.e., a finite density of both spin components) and there
are two gapless linear modes, describing excitations of the global
phase, and the relative phase of the two components.  When $|\Delta|$
becomes large enough to cause a transition to a circularly polarized
state, only a single gapless mode survives, describing phase modes of
the condensed component.  On the other hand, the presence of pumping
and decay is known to replace the linear dispersion of the phase modes
with a diffusive behavior at small
momentum\cite{szymanska06:prl,szymanska07,wouters07:opo,wouters07:bec}.
The aim of this section is to see how these two effects are combined.
The result is that while the global phase mode remains diffusive, the
real part of the relative phase mode can be either gapped, linearly
dispersing, or diffusive according to the value of $\Delta$ chosen.
In the following, we will first present numerical results for the
normal-mode frequencies, and then discuss how these can be
straightforwardly interpreted in the same Josephson regime as
discussed above.

\begin{figure}[htpb]
  \centering
  \includegraphics[width=3.2in]{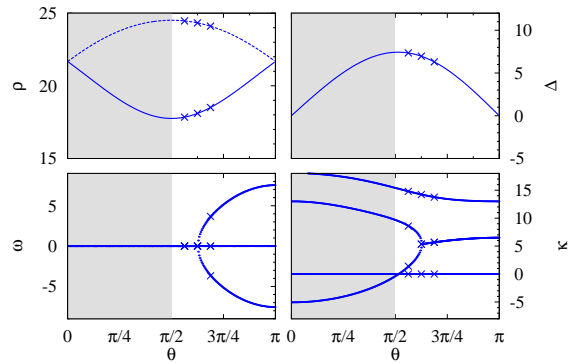}
  \caption{(Color online) Steady states, and damping of uniform fluctuations.  All
    panels are plotted as a function of the phase difference,
    $\theta$, between the two circular polarization components.  Top
    left: densities of components in steady state solution.  Top
    right: Detuning corresponding to given solution.  Bottom: Real
    (left) and imaginary (right) parts of frequency for small
    fluctuations about the steady state.  The gray shaded region is
    unstable to small fluctuations.  The crosses mark the conditions
    used for the finite $k$ spectra in
    Fig.~\ref{fig:uniform-stability}.  Plotted for $\alpha=6.5,
    \sigma=0.3, J=1.0, u_a=1.1$.}
  \label{fig:uniform-stability}
\end{figure}

\begin{figure}[htpb]
  \centering
  \includegraphics[width=3.2in]{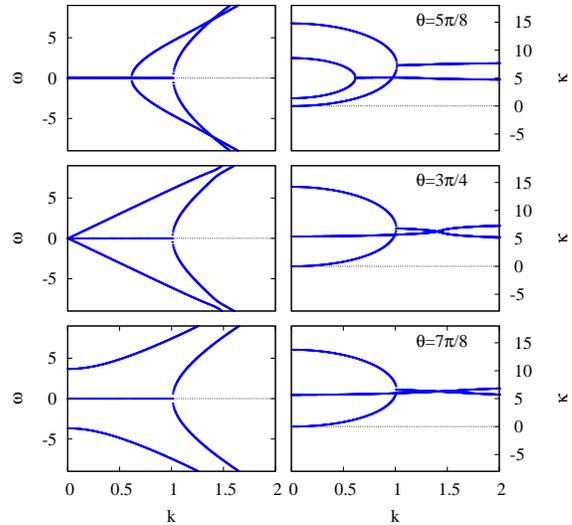}
  \caption{(Color online) Spectra of normal modes in a uniform system;
    showing real 
    (left) and imaginary (right) parts of each mode separately.  The
    three rows correspond to the three points marked by crosses in
    Fig.~\ref{fig:uniform-stability}.  In all cases the total density
    modes are diffusive, while the spin mode changes from diffusive
    (top row, $\theta$ nearest $\pi/2$), to linearly dispersive
    (middle row) to underdamped spin oscillations (bottom row,
    $\theta$ nearest $\pi$).}
  \label{fig:uniform-spectrum}
\end{figure}

For the purpose of numerically calculating the normal modes, it is
simplest to consider the Bogoliubov parametrization of fluctuations
at some wavevector $k$ (allowing for decay) as:
\begin{multline}
  \label{eq:8}
  \psi_{L,R}(t,r) = e^{-i\mu t} \psi^0_{L,R} + \\
  e^{-i\mu t}\left(
    u_{L,R} e^{-i\omega t - \kappa t - i \vect{k}\cdot\vect{r}} +
    v_{L,R}^\ast e^{i\omega t - \kappa t + i \vect{k}\cdot\vect{r}}
  \right).
\end{multline}
Substituting this into Eq.~(\ref{eq:2}), and linearizing in $u,v$
yields the secular equation $\Det [(\omega - i \kappa) \mathbbm{1} -
M] = 0$, where in the basis $\chi = ( u_L, v_L, u_R, v_R)^T$, the
matrix $M$ is
\begin{equation}
  \label{eq:9}
  M=
  \left(
    \begin{array}{cccc}
      A_L & B_L & C_L & D_L \\
      -B_L^\ast & -A_L^\ast  & -D_L^\ast & -C_L^\ast \\
      A_R & B_R & C_R & D_R \\
      -B_R^\ast & -A_R^\ast  & -D_R^\ast & -C_R^\ast 
    \end{array}
  \right).
\end{equation}
Noting that for plane waves, the kinetic energy term $-\nabla^2 \to
k^2$, the matrix elements are:
\begin{align*}
  A_L &= k^2 - J \frac{\psi_R^0}{\psi_L^0} + (1-i\sigma) |\psi_L^0|^2, &
  B_L &= (1 - i\sigma) (\psi^0_L)^2 \\
  C_L &= J + (1-u_a) \psi^0_L \psi_R^{0\ast}, &
  D_L &= (1-u_a) \psi^0_L \psi^0_R,
\end{align*}
and similarly with $L \leftrightarrow R$.  The normal modes calculated
this way are shown in
Figs.~\ref{fig:uniform-stability},~\ref{fig:uniform-spectrum}.
Figure~\ref{fig:uniform-stability} shows the modes at $k=0$ (lower
panels) as well as the value of $\Delta$, and the densities
$\rho_{L,R}$ (upper panels) corresponding to a solution with a given
value of $\theta$. [This is to make use of the fact that the steady
state of Eqs.~(\ref{eq:4}--\ref{eq:6}) can be found explicitly for
$\Delta, \rho_{L,R}$ as a function of $\theta$ (see
Ref.~\onlinecite{wouters08:prb}).]

In Fig.~\ref{fig:uniform-stability}, the range of $\theta$ is
restricted to $[0, \pi]$, since the range $[\pi, 2 \pi]$ is equivalent
to this, after swapping $L \leftrightarrow R$ and $\Delta \to -
\Delta$.  Within this range, only values $\theta > \theta_c \simeq
\pi/2$ correspond to stable solutions. As only the modes at $k=0$ are
shown, there is always a zero mode corresponding to global phase
rotations.  The other three modes divide into an overdamped mode
(largest imaginary part), and a pair of modes which can be either
overdamped or underdamped, as seen by the bifurcation of the real part
at $\theta = 1.964$.  Three values of $\theta$ (corresponding to three
different applied detunings $\Delta$) are chosen to illustrate the
overdamped, critically damped, and underdamped cases, and the normal
modes with finite $k$ are shown for these points in
Fig.~\ref{fig:uniform-spectrum}.  In the overdamped case (for $\theta$
near $\pi/2$), both the relative-phase and global-phase modes are
diffusive at small $k$. When underdamped, the relative-phase mode
always has a nonzero real frequency. When critically damped, the real
part of the relative-phase mode has a linear dispersion.  We will next
discuss how this behavior can be understood in the Josephson regime.

To understand the modes qualitatively, it is clearest to work in the
basis of $R, z, \theta, \phi$.  Extending
Eqs.~(\ref{eq:4}--\ref{eq:6}) to include spatial gradients via a
Madelung transformation, one finds the equations:
\begin{align}
  \dot{R} &= 
  -  \nabla \left( 2 R \nabla \phi + z \nabla \theta \right)
  +  2 \alpha R - 2 \sigma (R^2 + z^2) 
  \\
  \dot{z} &=
  -  \nabla \left( 2 z \nabla \phi + R \nabla \theta \right)
  + 2 (\alpha - 2 \sigma R) z
  \nonumber\\
  &\qquad- 2 J \sqrt{R^2 - z^2} \sin \theta
  \\
  \dot{\phi} &= \frac{X_L + X_R}{2}
  - (2-u_a) R - \frac{J R}{\sqrt{R^2 - z^2}} \cos(\theta) 
  \\
  \dot{\theta} &= (X_L-X_R) - 2 u_a z - \frac{ 2 J z}{\sqrt{R^2 - z^2}}
  \cos(\theta) - \Delta,
\end{align}
where we have introduced the shorthand $X_{L,R} = (\nabla^2
\sqrt{\rho_{L,R}})/\sqrt{\rho_{L,R}} - (\nabla \theta_{L,R})^2$, with
$\theta_{L,R} = \phi \pm \theta/2$.  When considering fluctuations
about the uniform state, the gradient terms simplify, as expressions
such as $(\nabla \theta)^2$ are necessarily second order in the
fluctuations, and so may be neglected.  We will make the same
approximations as led to Eq.~(\ref{eq:7}) (i.e., assume $J \ll
u_a R, z \ll R$), and use the steady-state solution in this limit ($R =
\alpha/\sigma, J \sin(\theta) = - \sigma z$) to simplify the equations
for fluctuations.  If one writes the fluctuations as $\chi = (\delta
R, \delta z, \delta \phi, \delta \theta)^T$ then since these are real
variables, the normal modes have the time dependence $\chi(t) = \chi
\exp(-i\omega t-\kappa t)$, giving a
secular equation:
\begin{equation}
  \label{eq:11}
  \Det \left[
    (i\omega+\kappa) \mathbbm{1} + M_0 + k^2 M_1 + \mathcal{O}(k^4)  
  \right] = 0,
\end{equation}
where $M_n$ indicates an expansion in powers of momentum, to
understand the small $k$ dispersion.  The first two terms are:
\begin{align}
  \label{eq:12}
  M_0 &= \left(
    \begin{array}{cccc}
      - 2 \alpha & 0 & 0 & 0 \\
      - 2 \sigma z & - 2 \alpha & 0 & - 2 J R \cos(\theta) \\
      -2 + u_a & 0 & 0 & 0 \\
      0 & -2 u_a & 0 &0
    \end{array}
  \right)
  \\
  M_1 &= \left(
    \begin{array}{cccc}
      0 & 0 & 2 R & z \\
      0 & 0 & 2 z & R \\
      -1/2R & z/2R^2 & 0 & 0 \\
      z/R^2 & -1/R & 0 & 0 
    \end{array}
  \right)
\end{align}

The nature of the $k=0$ normal modes is immediately clear: $\phi$
has no restoring force, and so has a zero frequency oscillation; $R$
has a damping rate $2 \alpha$, and so describes a decaying mode; $z$
and $\theta$ are mixed, and have modes with frequencies given by:
\begin{equation}
  \label{eq:13}
  (i \omega+\kappa)(i \omega + \kappa - 2 \alpha) - 4 u_a J R \cos(\theta) = 0.
\end{equation}
Noting that $\cos(\theta)$ is negative, and that the prefactor is the same
as that of $\sin(\theta)$ in Eq.~(\ref{eq:7}), one may write:
\begin{equation}
  \label{eq:10}
  - 4 u_a J R \cos(\theta) = \Omega_p^2,
\end{equation}
where $\Omega_p(\theta)$ is a ``plasma frequency'' describing the
restoring force as a function of angle.  The normal modes are $\omega
-i\kappa= - i \alpha \pm \sqrt{\Omega_p^2 - \alpha^2}$.  The
transition between under- and overdamping occurs because for $\theta
\simeq \pi$ the restoring force is large so $\Omega_p > \alpha$, while
as $\theta \to \theta_c \simeq \pi/2$, the restoring force vanishes,
and so there is an intermediate value of $\theta$ (and hence
  $\Delta$) at which $\Omega_p = \alpha$, describing the critical
damping.  Note that whatever the choice of parameters, an overdamped
regime will always exist since $\Omega_p$ must always tend to zero at
$\theta_c$. However if $\alpha$ is sufficiently large the underdamped
regime may vanish.  The approximate expressions for the eigenvalues
$\omega -i\kappa= 0, -i\alpha \pm \sqrt{\Omega_p(\theta)^2 -
  \alpha^2}, - 2 i \alpha$ match the general form seen in the bottom
panel of Fig.~\ref{fig:uniform-stability}, it is apparent that the
full problem has some additional variation of damping rate with
$\theta$, not captured by the approximations used in the above
expressions.

One may now explain the linear dispersion at this critical damping
as arising from degenerate perturbation theory, allowing a $k^2$
perturbation to give a $\propto k$ splitting.  Setting $\Omega_p = \alpha$,
and writing $\omega -i\kappa= - i \alpha + \eta$, then expanding Eq.~(\ref{eq:11}) to
leading order in $\eta, k$ gives:
\begin{equation}
  \label{eq:14}
  0 = \eta^2 - k^2 \left[ \frac{\alpha^2}{2 u_a R} + 2 u_a R \right],
\end{equation}
hence describing modes $\omega -i\kappa= -i\alpha + c_{\text{eff}} k$.
Note that although these modes have a linear dispersion of the real
part as a function of $k$, they have a lifetime that remains finite as
$k\to 0$.  As such, this may provide an example where linear
dispersion of a given mode in a condensed system need not imply
superfluidity of the associated density.  Further work is required to
determine the current-current response function in this system, which
will determine whether there is a difference of transverse and
longitudinal response for spin currents, however superfluidity of the
spin current is not expected here, since the spin orientation is
locked by the Josephson term $J$.  The linear dispersion arises only
when the damping matches frequency of this phase-locking term,
producing critical damping.


\section{Stability of cross-polarized vortices with pumping and decay}
\label{sec:stab-cross-polar}

When an harmonic trap is introduced, the existence of steady state
currents can destabilize the Thomas--Fermi like density profile,
and lead to spontaneous rotating vortex lattices\cite{keeling08:gpe}.
Before addressing how this instability interacts with the polarization
degree of freedom, we first discuss the simpler question of the
dynamics and stability of individual vortices of opposite
polarization.  One should note first that a single vortex in a pumped
decaying condensate already has a more complicated structure than the
vortex without pumping and decay --- the vortex becomes a spiral
vortex, with both radial and azimuthal currents; see
appendix~\ref{sec:spir-vort-single} for further discussion.

\begin{figure}
\includegraphics[width=3.2in]{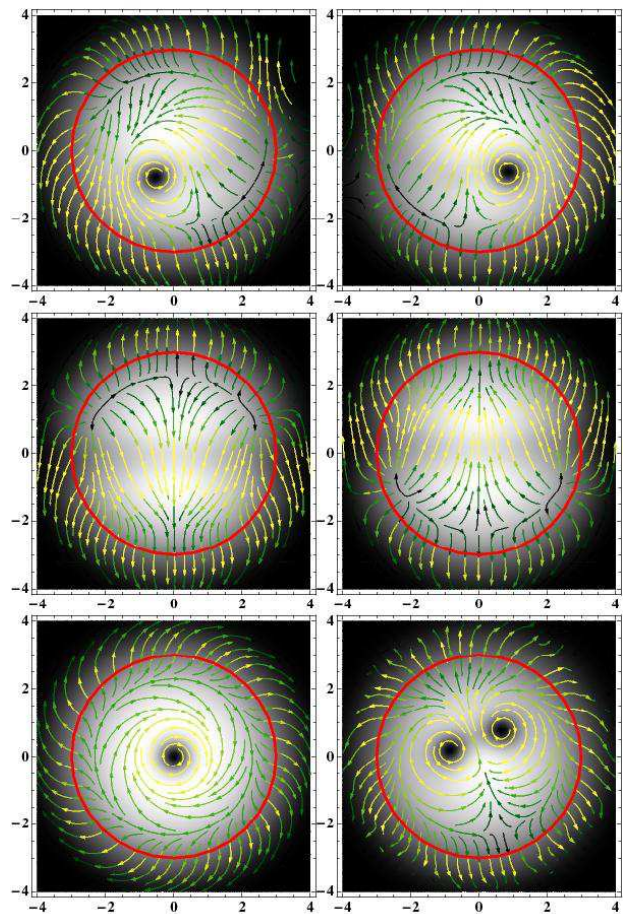}

\caption{(Color online) The outcome of instability of the vortex
  state $(+1,-1)$ for $\alpha=4.4, \sigma=0.3, r_0=3, \Delta=0,$ and
  $J=0.5$ (top row), $J=1$ (second row), $J=1.5$ (left bottom) and
  $J=2$ (right bottom). The initial state is
  $\psi_\pm=\sqrt{\Theta(\mu_{\text{TF}}-r^2)}(x\pm{\rm
    i}y)/\sqrt{r^2+1/(\mu_{\text{TF}} \delta)}$ where $\psi_R=\psi_-,
  \psi_L=\psi_+$, $\mu_{\text{TF}}=3\alpha/2\sigma$ and $\delta$ is the
  vortex core parameter; see Appendix~\ref{sec:spir-vort-single}.
  Depicted are the density
  plots  and streamlines of $\psi_L$ (left panels in first
  and second row, bottom row) and $\psi_R$ (right panels in first and
  second row). Luminosity is proportional to the magnitude of
  the density for
  density plots and to the velocity for streamlines.  The size
  of the pumping 
  is shown as a circle of radius $r_0=3$. For $J=0.5$ the final state
  consists of two precessing vortices of opposite circulation. For
  $J=1$ the system evolves into a pair of two gray solitons. For
  $J=1.5$ the vortex of the negative circulation in $\psi_R$ leaves
  and a vortex of positive circulation enters the cloud. For $J=2$
  both vortices leave the condensate and two pairs of vortices of
  opposite circulation enter. For $J=1.5$ and $J=2$ $\psi_L=\psi_R$,
  so only $\psi_L$ is shown. }
\label{jjj}
\end{figure}

\begin{figure}
\includegraphics[width=3.2in]{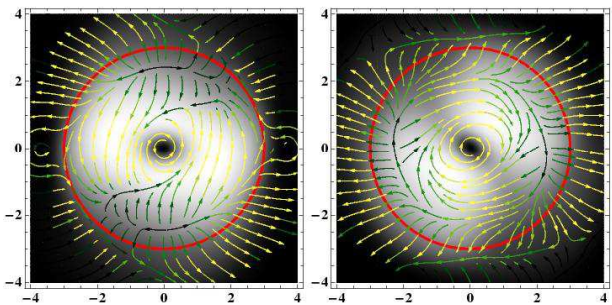}
\caption{(Color online) Stable state of cross-polarized vortices
    $(+1,-1)$ for $\alpha=4.4, \sigma=0.3, r_0=3, \Delta=8,$ and
    $J=1$. Depicted are the  density plots and streamlines
    of $\psi_L$ (left) and $\psi_R$ (right). Luminosity is
    proportional to the magnitude of the density for
    density plots, and 
    velocity is shown by streamlines. The size of the pumping is shown
    as a circle of radius $r_0=3$.}
\label{oppVort}
\end{figure}

Considering a spinor condensate system with vortices in both
polarization components, there is an interplay between the weak
attractive coupling that invites the vortices to have their cores
aligned irrespective of their circulation, the Josephson coupling that
discourages the alignment of vortices of opposite circulation and the
detuning that stabilizes such alignment. To illustrate these
possibilities we consider several examples. The notation $(\pm n, \pm m)$
refers to condensates with vortices of topological charge $\pm n$
($\pm m$) in left (right) condensate. From Eq.~(\ref{eq:2}), with
$v(r) = r^2$, the following set of stability scenarios are found:

\begin{description}
\item[$J=0$:]  All $(\pm 1,0)$ and $(\pm 1, \pm 1)$ vortex complexes are
dynamically stable.
\item[$J\ne 0, \Delta=0$:] Solutions $(+1,+1)$ are stable, $(\pm 1,0)$
and $(+1,-1)$ are unstable. Depending on the strength of $J$, vortices
may start precessing around the center of the trap, move beyond the
boundary forming either a pair of rarefaction waves with opposite
velocities or a complex $(+1,+1)$, or disappear at the condensate's edge
and bringing in two aligned pairs of vortices $(+1,+1)$ and
$(-1,-1)$. Fig.~\ref{jjj} illustrates these possibilities for
$\alpha=4.4, r_0=3, \sigma=0.3$ and $J=0.5,1,1.5,2$.

\item[$J\ne 0, \Delta\ne 0$:] For a given $J$, any sufficiently large
  $\Delta$ allows the vortex complexes $(+1,-1)$ and $(\pm 1,0)$ to
  stabilize. Fig.~\ref{oppVort} shows such stabilized $(+1,-1)$
  complex for $J=1, \Delta=8.$
\end{description}

\begin{figure}
\includegraphics[width=3.2in]{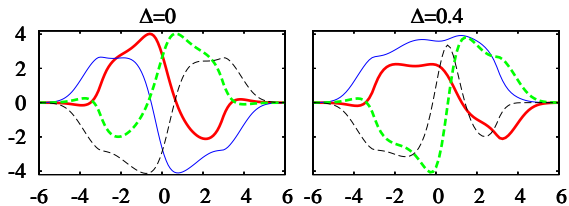}
\caption{(Color online) Real (thick lines) and imaginary (thin
  lines) parts
  of $\psi_L$ (solid lines) and $\psi_R$ (dashed lines) of the
  rarefaction wave complex as a stationary solution of
  Eq.~(\ref{eq:2}) with $v(r)=r^2$ for $\alpha=4.4, \sigma=0.3, J=1$
  and $\Delta=0$ (left) and $\Delta=0.4$ (right) plotted along the
  axis of density symmetry. The functions are antisymmetric with
  respect to the origin $\psi_L(0,-y)=\psi_R(0,y)$ for $\Delta=0$, but
  not for $\Delta\ne 0$.}
\label{grey}
\end{figure}

The stationary state shown in Fig.~\ref{jjj} for $J=1$ is the
 vorticity-free, but not radially
  symmetric, solution of
Eq.~(\ref{eq:2}). These are analogous to the rarefaction solitary
waves of the nonlinear Schr\"{o}dinger equation
discovered by Jones and Roberts 
\cite{jonesRoberts82}. In trapped 2D condensates these waves were also
found  \cite{mason08}: they form as two vortices
of opposite circulation disappear at the boundary of the condensate. In
the conservative GPE these waves propagate with velocities exceeding
the velocity of any vortex pair. In spinor damped/driven condensates
two  such solutions induce  flow in opposite
directions forming a stationary complex. The plots of the real and
imaginary parts of $\psi_L$ and $\psi_R$ shown in the left panel of
Fig.~\ref{grey}, and the density  plots of Fig.~\ref{jjj} can be
compared to the bottom panel of Fig.~9 and the
left panel of 
Fig.~10 of Ref.~\onlinecite{mason08}. It also follows from
simulations that $\psi_L(x,y)=\psi_R(x,-y)$, so the found stationary
state satisfies a one component Ginzburg--Landau equation
\begin{eqnarray}
i\partial_t \psi &=&[- \nabla^2 + r^2 + |\psi|^2\psi \nonumber \\
&& + {\rm i}(\alpha\Theta(r_0-r) - \sigma |\psi|^2)]\psi + J \psi(x,-y).
\label{gl_new}
\end{eqnarray}
This also suggests a way to generate solitary waves in spinor
condensates. The dark soliton (obtained by phase imprinting, for
instance) will undergo a transverse snake instability and form two
stationary pairs of vortices of opposite circulation, whereas starting
with a $(+1,-1)$ complex one will obtain stationary rarefaction
pulses.

\begin{figure}
\includegraphics[width=3.2in]{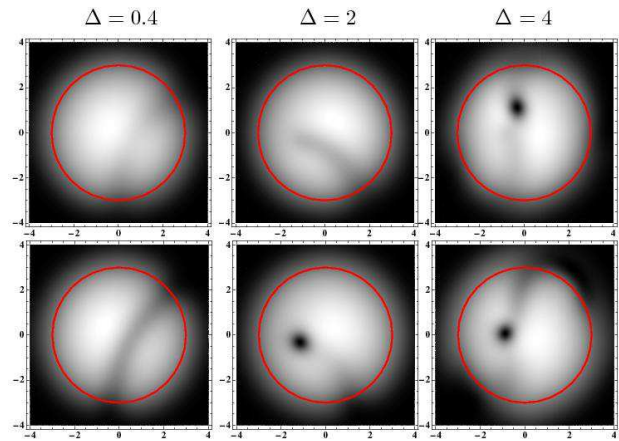}
\caption{(Color online) Density plots $|\psi_L|^2$ (top
    row) and $|\psi_R|^2$ (bottom row) for the outcome of instability
    of cross-polarized vortices obtained by numerical integration of
    Eq.~\ref{eq:2} with $v(r)=r^2$, $\alpha=4.4, \sigma=0.3, J=1$ and
    various $\Delta$. The initial state is the same as in
    Fig.~\ref{jjj}. Luminosity is proportional to density.  The size
    of the pumping is shown as a circle of radius $r_0=3$. For
    $\Delta=0.4$ two rarefaction pulses form a stationary complex. For
    $\Delta=2$ a vortex of negative circulation in the $R$
    component 
    is coupled to a rarefaction pulse in the $L$ component. For
    $\Delta=4$ the two vortices of opposite circulation do not
    align their 
    cores. For $\Delta=2$ and $\Delta=4$ the complexes precess around
    the center with the individual vortices moving along an epitrochoid;
    see also Fig.~\ref{motionVort}. }
\label{possibleDelta}
\end{figure}

As $\Delta$ increases, the rarefaction waves in two components lose
their antisymmetry and the density minima move away from the center
as the right panel of Fig.~\ref{grey} illustrates. For intermediate
$\Delta$, complexes combining a rarefaction wave in one component with
a precessing vortex in the other component emerge; for even larger
$\Delta$ two vortices of opposite circulation move around in the
condensate. Finally, for even larger $\Delta$, their cores overlap and
vortices move to the center of the condensate. These possibilities are
illustrated on Figs.~\ref{oppVort} and \ref{possibleDelta}.  The
cross-polarized vortices deform the condensate to a slightly oblate
form as seen in the density  plots of Fig.~\ref{oppVort}. The
streamlines and the density  plots indicate that the vortices
coexist with rarefaction waves that rotate in the counterclockwise
direction.

\begin{figure}
\centering
\includegraphics[width=1.6in]{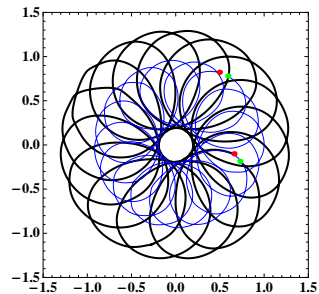}
\caption{(Color online) Trajectories of vortices in the complex
    $(+1,-1)$ obtained by numerical integration of Eq.~\ref{eq:2} with
    $v(r)=r^2$, $\alpha=4.4, \sigma=0.3, J=1$ and $\Delta=4$, and with
    pumping in a radius $r_0 = 3$.  These parameters are the same as
    for the density profiles in the right hand column of
    Fig.~\ref{possibleDelta}.  The trajectory of the vortex of
    positive (negative) circulation in the left (right) component is
    shown as thick black (thin blue) line. The start (end) points of
    the time interval plotted are shown as red (green) filled circles.
    Note that the vortex trajectories are not closed.}
\label{motionVort}
\end{figure}

 The vortex trajectories with pumping, decay and a spin degree
  of freedom are nontrivial.  For a one-component conservative GPE, a
  single vortex moves perpendicularly to the background density gradient
  due to the Magnus force\cite{rubinstein94,kivshar98} (the speed of
  the motion is however nonuniversal, and depends on the global
  condensate shape \cite{mason08}).  For a two-component conservative
  GPE with vortices in both components, there is an additional
  advection of each vortex by the flow pattern of the
  other component. 
  Including also pumping and decay, the trajectories of the vortices
  are yet more complicated, as illustrated in Fig.~\ref{motionVort}
  for a $(+1,-1)$ complex with $J=1, \Delta=4$.  Both vortices move along
  trajectories closely resembling epitrochoids, and the distance between
  the vortices varies quasi-periodically with time.  Similarly complicated
  cycloid trajectories of vortices are known for two-layer fluids with
  one vortex in each layer --- such behavior has been seen for example in
  models of tropical vortices\cite{khandekar75}.

\section{The trapped system}
\label{sec:trapped-system}

We study the time-dependent problem defined by Eq.~(\ref{eq:2}) with a
harmonic trapping potential $v(r)=r^2$ by numerical integration, using
a fourth-order, finite-difference approximation in space and a
fourth-order Runge--Kutta method in time. Pumping is restricted to a
circular spot of radius $r_0$ centered at the bottom of the trap, as
described at the end of section~\ref{sec:model-spin-polar}.  In all of
the following, we take the experimentally realistic
parameters\cite{keeling08:gpe,kasprzak06:nature,richard05:prb}
$\alpha=4.4$ and $\sigma=0.3$. With this choice of parameters, the
critical radius can be estimated from the Thomas--Fermi approximation
for the problem without polarization\cite{keeling08:gpe} as
$r_{0_c}\simeq 4.7$.  This section will present and discuss the
transition between synchronized and desynchronized states, and the
interplay with the instability to vortex lattice formation.
Section~\ref{sec:critical-delta} first addresses the value of $\Delta$
for which a transition occurs, as one changes the size of the pumping
spot, and coupling $J$.  Section~\ref{sec:attractors} will then
discuss in greater detail the nature of the new attractors near the
critical $\Delta$ that occur when one includes spatial degrees of
freedom.

\subsection{Critical $\Delta$, bistability and internal Josephson
  effect}
\label{sec:critical-delta}

\begin{figure}[htpb]
  \centering
  \includegraphics[width=3.2in]{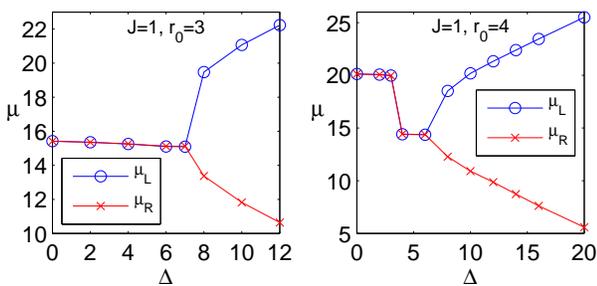}
  \caption{(Color online) Chemical potentials of the two polarization
  components 
  plotted against stepwise increased detuning $\Delta$. For solutions
  with time-dependent chemical potentials, a time average is shown.}
  \label{fig:chempotj1r3-j1r4}
\end{figure}

\begin{figure}[htpb]
  \centering
  \includegraphics[width=3.2in]{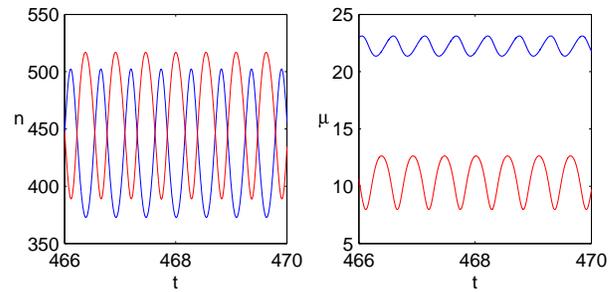}
  \caption{(Color online) Josephson oscillations in the desynchronized solution at
    $\Delta=12.0, r_0=3.0, J=1.0$ in the limit of large time. Left:
    $n=\int|\psi|^2\,d^2r$ as a function of time. Right: The chemical
    potentials of the two polarization components. (Time given in
    units of $2/\omega_0$. Blue and red color denote $L$ and $R$
    components, respectively.)}
  \label{fig:acjosephson}
\end{figure}

Let us start with the simplest trapped problem, taking $J=1.0$ and
$r_0=3.0$. We study the behavior of the system as the detuning
$\Delta$ is increased in the following way: starting at $\Delta=0$
from a Gaussian initial state, we let the system evolve until a steady
solution is reached. We then increase $\Delta$ in steps, taking for
each new $\Delta$ the final state at the previous $\Delta$ as the new
initial state. For each value of $\Delta$, the chemical potentials of
the two polarization components are calculated (see
appendix~\ref{app:numerical-precession} for details). The result is
shown in the left panel of Fig.~\ref{fig:chempotj1r3-j1r4}. The
components stay synchronized, sharing a common chemical potential, up
to $\Delta \simeq 7.0$. As the detuning is increased further, the components
desynchronize and AC Josephson oscillations occur between them
(Fig.~\ref{fig:acjosephson}).  Note that as the components
desynchronize, the chemical potentials show an oscillatory
time dependence. Fig.~\ref{fig:chempotj1r3-j1r4} shows a time average
in this case.

\begin{figure}[tpb]
  \centering
  \includegraphics[width=3.2in]{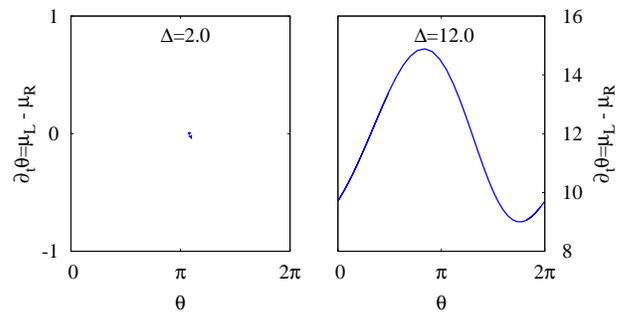}
  \caption{(Color online) Phase portraits for the trapped problem with
  $J=1.0$ and 
  $r_0=3.0$. Left: synchronized solution for small detuning. Right:
  desynchronized solution in the limit of large detuning (corrected
  for numerical precession).}
  \label{fig:r3progression}
\end{figure}

In terms of a suitably defined, spatially averaged  phase
difference $\theta$ between the components (see
  appendix~\ref{app:numerical-precession}), the desynchronization transition
corresponds to a transition from a fixed point to a limit cycle in the
$(\theta,\dot\theta)$ plane, in direct analogy to the two-mode
problem (see Fig.~\ref{fig:r3progression} and compare with
Fig.~\ref{fig:attraction-basin}).

\begin{figure}[htpb]
  \centering
  \includegraphics[width=3.2in]{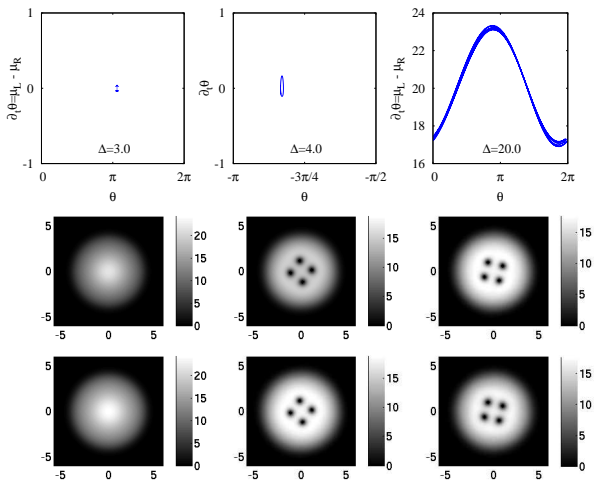}
  \caption{(Color online) Progression from fully synchronized solution
  via formation 
  of vortex lattice to fully desynchronized solution at $r_0=4.0$ and
  $J=1.0$. Top: phase portraits for the phase difference
  $\theta$ (a numerical precession has been removed from the
  $\Delta=20.0$ plot). Middle and bottom: total densities of the $L$ and $R$
  components, respectively.}
  \label{fig:r4progression}
\end{figure}

The right panel of Fig.~\ref{fig:chempotj1r3-j1r4} shows the chemical
potentials of the two polarization components as the detuning is
increased stepwise in a system pumped in a spot with radius
$r_0=4.0$. The progression from the synchronized solution at small
detunings to the desynchronized solution in the limit of large
detunings is now more complicated than for a small pumping
spot. $r_0=4.0$ is too small a pumping radius for vortices to form
spontaneously in the corresponding one-component
system\cite{keeling08:gpe}. Accordingly, the spinor system with zero
detuning develops circularly symmetric densities, identical in both
components (but with different phases for the two components). For
small enough detunings, the system stays synchronized, the densities
adjusting to accommodate the common, constant chemical
potential (left-most panels of Fig.~\ref{fig:r4progression}).

As the detuning increases beyond $\Delta\simeq 3.0$ the rotational
instability of the high-angular-momentum modes~\cite{keeling08:gpe}
reappears, and a four-vortex lattice forms in both components. At this
point, the smaller component is sufficiently depressed by the detuning
that the effective critical radius is smaller than the pumping
radius. The suppressed component becomes rotationally unstable and
drags the larger component along. This transition to a vortex-lattice
solution causes the (common) chemical potential to drop, as shown in
Fig.~\ref{fig:chempotj1r3-j1r4}.

While the two components remain on average phase locked, the drop in
chemical potential at the formation of the vortex lattice is
accompanied by a time-dependence of the chemical potentials, which
oscillate together around a common mean. Due to the amplitude of the
oscillations being slightly larger in the larger component, the fixed
point in the $(\theta,\dot\theta)$ plane turns into a small limit
cycle, but with only small variations of $\theta$ unlike the $2\pi$
periodic cycles in the desynchronized phase. (Middle panels of
Fig.~\ref{fig:r4progression}.)

Finally, as $\Delta$ becomes large, the components desynchronize
completely. This situation is shown in the right-hand panels of
Fig.~\ref{fig:r4progression}. The  difference in space-averaged
phase now traces out a limit cycle winding through the full $2\pi$.

\begin{figure}[htpb]
  \centering
  \includegraphics[width=3.2in]{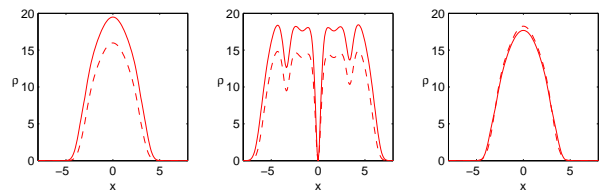}
  \caption{(Color online) Density profiles along the $x$-axis for representative
  examples of synchronized and desynchronized solutions. Left:
  synchronized solution without vortices ($r_0=3.0,
  \Delta=4.0$). Middle: synchronized solution with vortex lattice
  ($r_0=6.0, \Delta=4.0$). Right: desynchronized solution without
  vortices ($r_0=3.0, \Delta=12.0$). Dashed and solid lines indicate
  left and right-polarized components, respectively. $J=1.0$ for all
  three examples.}
  \label{fig:densityprof}
\end{figure}

The overall progression from synchronized solutions at small $\Delta$
to desynchronized solutions with Josephson oscillations holds very
generally for different values of the Josephson coupling $J$ and also
for both small (no vortices) and large (vortex lattice) pumping-spot
sizes. The mechanism in both cases is the same: the synchronized
solution is upheld by adjusting the densities and a steady
interconversion current forms. After desynchronization, the densities
revert to profiles as for the single component condensate, with the
average density set by the balance of pumping and decay, but with an
additional time-dependent interconversion
current. Fig.~\ref{fig:densityprof} shows density profiles of typical
synchronized solutions with and without vortices, and also a snapshot
of a desynchronized solution.

\begin{figure}[htpb]
  \centering
  \includegraphics[width=3.2in]{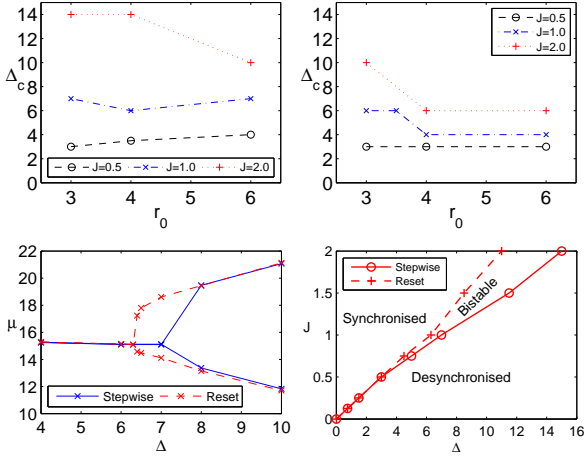}
  \caption{(Color online) Top: Highest $\Delta$ yielding synchronized solutions
    plotted as a function of $r_0$ for $J=0.5, 1.0$ and $2.0$. Left:
    $\Delta$ increased stepwise. Right: resetting initial conditions
    for each $\Delta$. Bottom left: bifurcation of chemical potentials
    (plotted as in Fig.~\ref{fig:chempotj1r3-j1r4}) for the two
    approaches, indicating the region of bistability for the example
    $J=1.0, r_0=3.0$. Bottom right: sketch of regions of stability of
    synchronized and desynchronized solutions for $r_0=3.0$. Compare
    with inset of Fig.~\ref{fig:ft-two-mode}. }
  \label{fig:deltac}
\end{figure}

We also performed calculations where we reset the initial
conditions to a Gaussian density profile, with equal phase of the two
polarization components for each value of $\Delta$.  Based on the
results of the two-mode model shown in
Fig.~\ref{fig:attraction-basin}, such initial conditions would be
expected to find the limit cycle whenever it exists, whereas the
stepwise increase of $\Delta$ discussed above is intended to follow
the fixed point.  In these calculations with resetting of initial
conditions, desynchronized solutions generally develop at smaller
$\Delta$, as one should expect if there is a region of coexistence of
synchronized and desynchronized solutions.  Fig.~\ref{fig:deltac}
shows the highest $\Delta$ yielding synchronized solutions for
different $r_0$ and different $J$, both for stepwise increased
$\Delta$ (top left) and calculation directly from Gaussian initial
conditions (top right). Comparing the two plots gives an estimate of the
region of coexistence. An example ($J=1.0, r_0=3.0$) showing the
region of bistability is given in the bottom left panel. Note that the
difference between stepwise increased $\Delta$ and reset to Gaussian
initial condition for each $\Delta$ vanishes for large detunings. This
need not be the case for larger $r_0$ or larger $J$, where more than
one desynchronized, meta-stable state (distinguished by numbers of
vortices) may be possible at high $\Delta$.

\begin{figure}[tpb]
  \centering
  \includegraphics[width=3.2in]{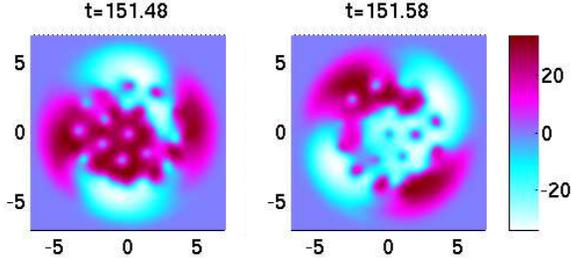}
  \caption{(Color online) Spatially nonuniform interconversion rate
  $\left.\partial_t\rho_1\right|_J$ at two different, but close,
  times. $J=1.0, r_0=6.0, \Delta=16.0$.}
  \label{fig:interconvmap}
\end{figure}

Our calculations suggest that the spatially extended system allows for
a rich variation of behaviors. Notably, interconversion due to the
Josephson coupling between the components need not be uniform in
space. The interconversion rate is easily calculated from the
time-dependent local phase
difference $\theta(\mathbf{r})=\phi_R(\mathbf{r})-\phi_L(\mathbf{r})$ as
\begin{equation}
\label{eq:drho1}
   \left.\frac{\partial\rho_L(\mathbf{r})}{\partial t}\right|_J 
   = -\left.\frac{\partial\rho_R(\mathbf{r})}{\partial t}\right|_J 
   =2J\sqrt{\rho_L(\mathbf{r})\rho_R(\mathbf{r})}
   \sin\left(\theta(\mathbf{r})\right),
\end{equation}
familiar from the Josephson effect as found in
any textbook\cite{tilley90,leggett01}. Note that Eq.~(\ref{eq:5})
describes the 
same physics for the two-mode model. Fig.~\ref{fig:interconvmap} shows
a map of 
the interconversion rate $\left.\partial_t\rho_1\right|_J$ at two
close points in time in the
large-time limit of the solution directly from Gaussian initial
conditions with 
$J=1.0, r_0=6.0, \Delta=16.0$. This solution is desynchronized and
exhibits Josephson oscillations. However, at any given time there is
interconversion in both directions at different points in space.

\begin{figure}[htpb]
  \centering
  \includegraphics[width=3.2in]{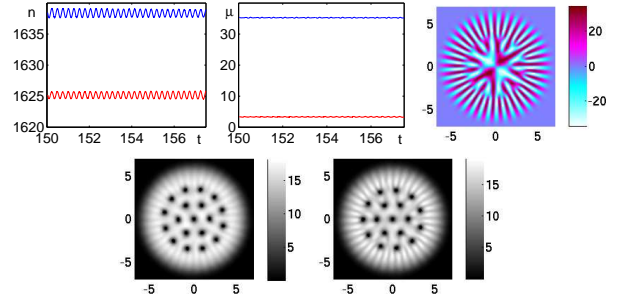}
  \caption{(Color online) Top from left to right: scaled mass $n=\int|\psi|^2\,d^2r$,
    chemical potentials and interconversion rate
    $\left.\partial_t\rho_1\right|_J$ for a desynchronized solution
    with counter-rotating components. Bottom: densities of the $L$ and
    $R$ polarization components, respectively. (Note that the gray
    scale has been adjusted for each component separately to
    accentuate the density modulations.) $J=1.0, r_0=6.0,
    \Delta=32.0$.}
  \label{fig:counter-rot}
\end{figure}

Calculations directly from Gaussian initial conditions at large
detuning with a large pumping spot ($r_0=6.0$) indicate that solutions
are possible in which the polarization components develop
counter-rotating vortex lattices. This leads to rapid density
modulations, particularly around the edge of the cloud, and a
corresponding pattern of interconversion in opposite directions. In
this case, the Josephson oscillations are suppressed and the
total (scaled) mass $n=\int|\psi|^2\,d^2r$ exhibits rapid,
small-amplitude oscillations. These effects are shown in
Fig.~\ref{fig:counter-rot}.

\subsection{Phase portraits of more complicated attractors}
\label{sec:attractors}

The spatial degree of freedom allows a number of behaviors of the
$(\theta,\dot\theta)$ phase portrait that are not possible in the
two-mode problem, which only allows fixed points (synchronized
solutions) and limit cycles with winding number $1$ (desynchronized
solutions). In the spatially extended system, these behaviors are
exemplified in Fig.~\ref{fig:r3progression} for a system without
vortices. Both these two classes of attractor can also be seen when a
vortex lattice exists. In addition, the spatial degree of freedom
gives rise to several new classes.

\begin{figure}[htpb]
  \centering
  \includegraphics[width=3.2in]{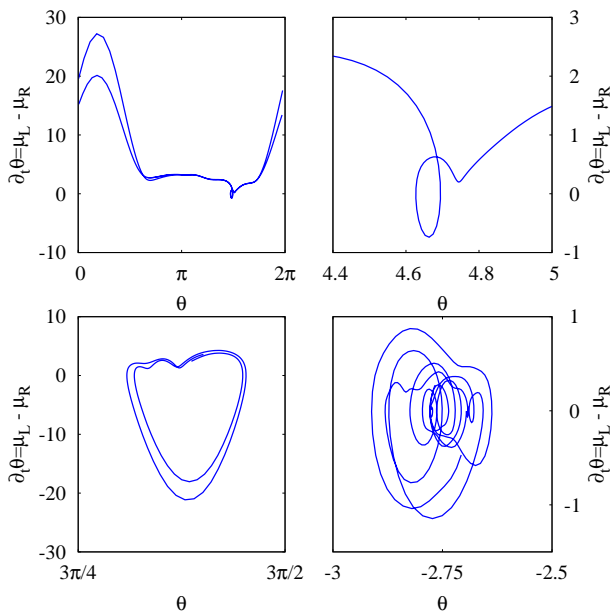}
  \caption{(Color online) Top: Limit cycle with winding number $2$ and
    retrograde 
    loop. Left panel shows a blow-up of the loop. A numerical
    precession has been removed from these plots. ($J=1.0, r_0=3.0,
    \Delta=6.4$.) Bottom left: Possibly quasi-periodic behavior
    similar to a limit cycle with winding number $0$ ($J=2.0, r_0=3.0,
    \Delta=14.0$ from stepwise increase). Bottom right: Possible
    chaotic attractor ($J=2.0, r_0=3.0, \Delta=10.0$ from stepwise
    increase).}
  \label{fig:phaseportraitzoo}
\end{figure}

We have already noted in Fig.~\ref{fig:r4progression} an example of a
synchronized limit cycle.  This can be distinguished from the
desynchronized limit cycle by the winding number, $\int \dot{\theta}
dt/2 \pi$, over one period.  The synchronized limit cycle has winding
number $0$, and the desynchronized cycles have winding number
  $1$.  We also find an example of limit cycle with winding number
$2$, shown in Fig.~\ref{fig:phaseportraitzoo}. This solution also
exhibits another behavior not possible in the two-mode model: a
retrograde loop. This solution appears in calculation directly from a
Gaussian initial condition at $\Delta=6.4$, which is barely above
$\Delta_c$ for $J=1.0, r_0=3.0$. For larger $\Delta$ the loop quickly
becomes a cusp and then disappears.

The bottom panels of Fig.~\ref{fig:phaseportraitzoo} show two other
behaviors found when stepwise increasing the detuning in a system
with a strong Josephson coupling ($J=2.0$). Both these solutions are
basically synchronized (the time average of the chemical potential is
the same for both components). However, the large detuning causes the
chemical potentials to differ at most instances in time, resulting in
behaviors similar to the limit cycles with winding number $0$, but
which appear to have a chaotic attractor and/or quasi-periodic
behavior.

\section{Experimental signatures}
\label{sec:exper-sign}

To directly observe rotating vortex lattices in experiments would
require time-resolved measurements on timescales of the order of the
trap frequency, which would be challenging with current experimental
configurations.  It is however possible to see signatures of a vortex
lattice in the momentum- and energy-resolved photoluminescence spectrum,
which can be directly measured in the far field.  The spectral weight
is given by the modulus squared of the Fourier transform of the
wavefunction:
\begin{equation}
\label{eq:spectrum}
 I(\omega,\mathbf{k}) =
   \left.\left|\int d^2r\,e^{-i\mathbf{k}\cdot\mathbf{r}}
   \int dt\,e^{-i\omega t}\psi(\mathbf{r},t)\right|\right.^2.
\end{equation}

\begin{figure}[htpb]
  \centering
  \includegraphics[width=3.2in]{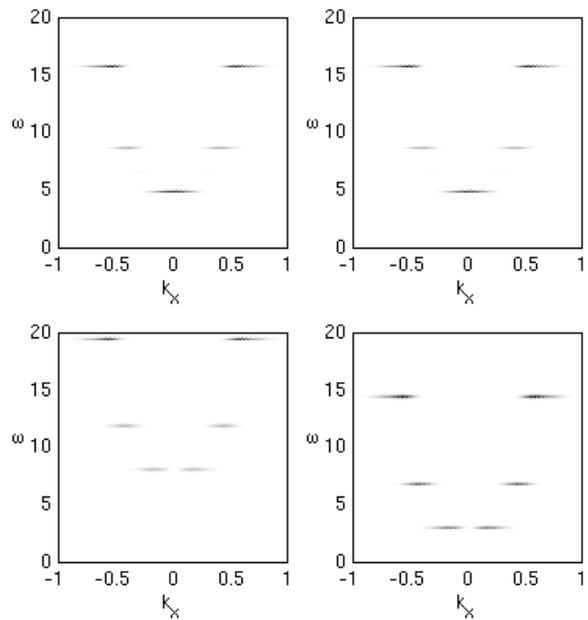}
  \caption{Spectral weight at $k_y=0$ for two vortex-lattice
  solutions. Left and right panels show left and right polarized
  components, respectively. Top: synchronized solution without a central
  vortex. ($J=0.5, \Delta=3.5, r_0=6.0$.) Bottom: desynchronized
  solution with a central vortex. ($J=0.5, \Delta=16.0, r_0=6.0$.)}
  \label{fig:spectra}
\end{figure}

As an illustration of this, Fig.~\ref{fig:spectra} shows the spectral
weight as a function of $(\omega, k_x, k_y=0)$.  The vortex lattice as
well as the desynchronization transition can be seen in the dispersion
spectrum. Fig.~\ref{fig:spectra} shows the spectra of the two
polarization components for two different solutions. The top panels
show a synchronized solution that exhibits a vortex lattice without a
central vortex. Each ring of vortices in the lattices shows up
as a side band in the spectrum. If there were no vortices, only the
bottom band would be present. As a contrast, the bottom panels show a
desynchronized solution that has a vortex lattice with a central
vortex. The presence of the central vortex means that the spectral
weight vanishes at $k_x=k_y=0$. Note also that the desynchronization
causes the spectra of the two components to be shifted relative to
each other, whereas in the synchronized case, the two spectra are
identical.

\section{Conclusions}
\label{sec:conclusions}

The interplay between the effects of pumping and decay in setting the
density profile, and the dynamics of the polarization degree of
freedom lead to a rich variety of possible nonequilibrium steady
states, as well as dynamical attractors for the polarized polariton
condensate in a harmonic trap.  Even neglecting spatial currents, the
two-mode model shows nontrivial behavior as a function of applied
magnetic field: with strong damping there is a simple transition
between synchronized and desynchronized states, whereas for weak
damping, a region of bistability exists, in which the state
established depends on the initial conditions.

Allowing for spatial fluctuations in a homogeneous system, the
plane-wave modes on top of the synchronized solution show a
second class of 
transition: if the phase locking term for the spin degree of freedom
is large enough, then at weak magnetic fields, this term will provide
sufficient restoring force for spin fluctuations to give underdamped
global spin oscillations.  As the magnetic field increases, the
restoring force for such spin waves decreases, eventually vanishing
when the synchronized solution becomes unstable.  Before this
instability occurs, there is a transition between underdamped and
overdamped spin oscillations, and at this transition, the spin wave
energies have a linear dispersion vs momentum.

Introducing vortices, the phase locking term favors co-alignment of
vortices in the two spin polarizations, but at sufficiently large
$\Delta$ anti-aligned vortex pairs may become stable. The
phase-locking term also makes it possible to obtain a stationary
complex of rarefaction waves. The experimental realization of such
solitary vorticity-free waves would demonstrate another aspect of
superfluid behavior\cite{kbnature,carusotto08} in the
incoherently pumped polariton system.

Considering the transition between synchronized and desynchronized
states in the inhomogeneous system, for sufficiently small pump spots,
the transition is similar to the two-mode model.  For larger pump
spots however, an instability to vortex-lattice formation may preempt
desynchronization, driven by the density imbalance required to sustain
a synchronized solution.  For both large and small pumping-spot
radius, the phase portraits near the critical detuning can be more
complicated, showing synchronized limit cycles with winding number 0,
desynchronized limit cycles with winding number 2, and chaotic
behavior.

In conclusion, the results presented here illustrate the wide variety
of dynamical behavior that can arise in spinor polariton condensates,
and suggest that experimental efforts to investigate such behavior
should be feasible.  

\acknowledgments{ J.~K.\ acknowledges discussions with P.~R.~Eastham,
  and funding under EPSRC grant no EP/G004714/1. M.~B.\
  acknowledges financial support from the Swedish Research Council.
  N.~G.~B.\ is grateful to Dr.~Natalia Janson for a useful discussion,
  and to the Isaac Newton Trust for financial support.}

\appendix

\section{Spiral vortices in the single component condensate}
\label{sec:spir-vort-single}

\begin{figure}
\includegraphics[width=3.2in]{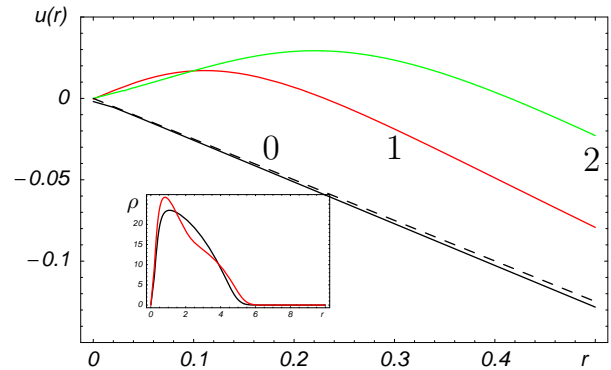}
\caption{(Color online) Radial velocity, $u(r)=\phi^{\prime}(r)$, of
  the vortex solutions of Eq.~(\ref{eq:GPE_rescale}) for $\alpha=1,
  \sigma=0.01$. The numbers next to the lines indicate the
  winding numbers 
  of the vortices (1 and 2) and the ground state (0). The
  TF
  approximation of the gradient flow of the ground state for small
  $\alpha$  and $\sigma$  $u(r)=-r
  \sigma(\mu -r^2)/6$, is given by the dashed line.  The inset
  compares the density, $\rho(r)$, of a vortex solution for $\alpha=4,
  \sigma=0.27$ ($\mu=30.28$, gray (red) line) with the density of the
  GPE vortex ($\alpha=0, \sigma=0$) with the same number of particles
  ($2 \pi\int  |\psi|^2\, r dr=1000$, $\mu=25.68$, black line). Both vortex
  solutions have topological charge $1$.}
\label{vortices}
\end{figure}

This appendix will discuss spiral vortices in the one-component version
of Eq.~(\ref{eq:2}), i.e., neglecting $J$ and $\Delta$.
It was shown in Ref.~\onlinecite{keeling08:gpe} that below the critical
radius of pumping for the formation of a vortex lattice there exist
states with one or a few spiral vortices.
 By combining vorticity with pumping and decay, one has both radial and
azimuthal supercurrents, both of which modify the density profile, and
so these currents interact.
This is shown in Fig.~\ref{vortices}, which shows both the radial
velocity (main figure) for vortex solutions, and the density profile
(inset).
The vortex solution requires vanishing density at the origin,
this makes the region at small radii a region  of net gain,
and leads to an outward-flowing current.
Thus, in the vortex solutions, there exists both a local maximum of
radial current, and a point at finite radius (of the order of the
healing length) at which this current vanishes.

In polar coordinates $(r,\theta)$ the wavefunction of a spiral vortex
of topological charge $s$ in the one-component polariton condensate
takes the form $\psi_v=f(r)\exp[i \phi(r)+s \theta]$, with the
equations governing $f$ and $u(r)=\phi'(r)$ given by
\begin{equation}
\label{eq:GPE_rescale}
\begin{split}
&\frac{d}{rdr}\Biggl(rf^2u\Biggr)=(\alpha\Theta(r_0-r)-\sigma f^2)f^2,\\
&f''+\frac{1}{r}f' + \Biggl(\mu - u^2
-\frac{s^2}{r^2}-f^2-r^2\Biggr)f=0.
\end{split}
\end{equation}
Around the center of the trap $f(r)$ and $u(r)$ can be found
recursively in the form of the power series  
\begin{equation}
f(r)=\sum_{i=1}^\infty a_i r^{|s|(2i-1)}, \quad
u(r)=\sum_{i=1}^\infty b_i r^{2i-1}.
\label{sys}
\end{equation}
To the leading order $u(r)\sim \alpha/2(s+1) r$ showing that the
stronger the pumping the larger the outward velocity is in the
vortex core. Only vortices of topological charge $1$ are dynamically
stable. For these, $a_1 \approx \delta \mu, a_3=-a_1 \mu/8 ...,
b_2=(\alpha\mu-8\sigma a_1^2)/48 ...$, where the numerical value
of the vortex core parameter $\delta\approx 0.583$ was first
calculated by Pitaevskii \cite{pitaevskii61} for the straight line
vortex of uniform Gross--Pitaevskii condensate.

\section{Irregular vortex lattices}
\label{app:irregular-lattices}

\begin{figure}[htpb]
  \centering
  \includegraphics[width=3.2in]{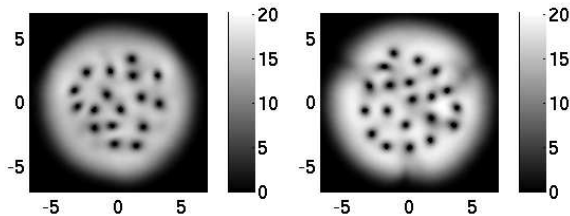}
  \caption{Density of the left- and right-polarized components,
  respectively, showing an irregular vortex lattice.}
  \label{fig:irregular-density}
\end{figure}

We note that around and above critical detuning, solutions to
Eq.~(\ref{eq:2}) may exhibit irregular density profiles and irregular
vortex lattices. This happens most frequently when solutions are found
starting from Gaussian initial conditions, but it may also happen for
large $r_0$ when $\Delta$ is increased stepwise. As an example, we
show a solution obtained from Gaussian initial conditions at $J=1.0,
r_0=6.0, \Delta=6.0$ in Fig.~\ref{fig:irregular-density}. This
solution shows an irregular vortex lattice, reminiscent of turbulent
behavior. 

\section{Chemical potentials, phase portraits and numerical precession}
\label{app:numerical-precession}

In order to numerically find the chemical potential of a condensate
with a rotating vortex lattice, it is useful to use:
\begin{equation}
  \label{eq:15}
  \mu - 2 s \Omega = \frac{
    \int  e^{i s \vartheta} i \partial_t \psi(\vect{r}) d^2r}{
    \int  e^{i s \vartheta} \psi(\vect{r}) d^2r}
\end{equation}
along with the equation:
\begin{equation}
  \label{eq:16}
  \mu - 2 \langle L_z \rangle \Omega = \frac{%
    \int  \psi^\ast(\vect{r}) i \partial_t \psi(\vect{r}) d^2r}{%
    \int   |\psi(\vect{r})|^2 d^2r}.
\end{equation}
In many cases, Eq.~(\ref{eq:15}) with $s=0$ is sufficient to determine
$\mu$. However, when there is a vortex at the center of the cloud, the
numerator and denominator both vanish, whereas one of $s=\pm 1$ will
give a well defined value.  In such cases, it is then necessary to use
Eq.~(\ref{eq:16}) and the integral for $\langle L_z \rangle$ to eliminate
$\Omega$.

For plotting the phase portraits, it is necessary also to calculate
the associated phase, $\phi$ such that $\dot{\phi} = \mu$.  This can
similarly be defined as:
\begin{equation}
\label{eq:average-phase}
   \phi \equiv
   {\rm Im}\left\{\ln\left[\int\psi(\mathbf{r})\,d^2r\right]\right\}.
\end{equation}
However, when there is a central vortex, the same difficulty will
arise.  To avoid this problem, two possible approaches are used.  In
some most cases it is sufficient to calculate $\dot{\theta} =
\mu_L-\mu_R$, and numerically integrate this equation to find
$\theta(t)$.  Numerical errors in evaluating $\mu_{L,R}$ can however
introduce unphysical precession into the plot of $\theta,
\dot{\theta}$.  To avoid this, in those cases where the nature of the
portrait is simple, we use $\theta_{i+1}=\theta_i+\dot\theta\times
dt\times f$, with $f\simeq 1$ adjusted to remove the spurious
precession.  This method has been used in
Figs.~\ref{fig:r3progression}, \ref{fig:r4progression} (right
  hand panel) and \ref{fig:phaseportraitzoo}.

Where the phase portrait involves finer structure, such as the
middle panel of
Fig.~\ref{fig:r4progression}, the phase portrait has instead been
obtained by:
\begin{equation}
  \label{eq:17}
  \begin{split}
  \theta = 
  {\rm Im}&\left\{
    \ln\left[\int e^{i s \vartheta} \psi_R(\mathbf{r})\,d^2r\right]\right.\\
    &-
    \left.\ln\left[\int e^{i s \vartheta} \psi_L(\mathbf{r})\,d^2r\right]
  \right\},
  \end{split}
\end{equation}
with $s=0,\pm 1$ according to whether there is a central vortex.  Since the
two components are corotating in this case, such a definition satisfies:
\begin{equation}
  \label{eq:18}
  \dot{\theta} = \left( \mu_L - 2 s \Omega \right) - 
  \left( \mu_R - 2 s \Omega \right) = \mu_L - \mu_R
\end{equation}
as required.

\end{document}